\documentclass[aps,preprint]{revtex4}

\def\be{\begin{eqnarray}}
\def\ee{\end{eqnarray}}

\begin{document}
\title{Magic Neutrino Mass Matrix and the Bjorken-Harrison-Scott Parameterization}
\author{C.S. Lam}
\address{Department of Physics, McGill University,\\ Montreal, QC, Canada H3A 2T8\\
and\\
Department of Physics and Astronomy, University of British Columbia,  Vancouver, BC, Canada V6T 1Z1 \\
\email: Lam@physics.McGill.ca}

\begin{abstract}
Observed neutrino mixing can be described by a tribimaximal MNS matrix. The resulting neutrino mass matrix in the 
basis of a diagonal charged lepton mass matrix is both 2-3 symmetric and magic. By a magic matrix, 
I mean one whose row sums and column sums are all identical. I study what happens if
2-3 symmetry is broken but the magic symmetry is kept intact. In that case, the mixing matrix 
is parameterized by a single complex parameter $U_{e3}$,
in a form discussed recently by Bjorken, Harrison, and Scott.
\end{abstract}
\maketitle

\section{Introduction}
A global fit on neutrino oscillation experiments has established the mixing parameters to be \cite{FLMP} 
$\sin^2\theta_{13}=0.9{+2.3 \atop -0.9}\times 10^{-2}$, $\sin^2\theta_{12}=0.314(1 {+0.18\atop -0.15})$,
and $\sin^2\theta_{23}=0.44(1 {+0.41\atop -0.22})$, all at 95\% C.L.    Squared mass differences
are also known but neither the Dirac phase nor the two Majorana phases has been measured. The mixing angles are
consistent with a tribimaximal structure first proposed by Harrision, Perkins, and Scott (HPS)
\cite{HPS}, whose mixing matrix is shown in eq.~(1). The corresponding mixing parameters are
$\sin^2\theta_{13}=0$, $\sin^2\theta_{12}=0.333$, and $\sin^2\theta_{23}=0.50$.
Many dynamical models have been proposed to explain the mixing  \cite{MODELS}, usually through a horizontal group
spontaneously broken by appropriately chosen scalar fields and parameters.
Often these models also predict some deviations from the HPS mixing.
We shall not discuss such models, but shall persue possible deviations 
through a symmetry structure of the neutrino mass matrix.

The HPS mixing possesses two remarkable properties. In the basis where the charged lepton mass matrix is diagonal,
it gives rise to a neutrino mass matrix that is 2-3 symmetric, so that when we interchange the second and the third
rows, and simultaneously the second and the third columns, the mass matrix remains the same. 
See eq.~(2) below. Such a
2-3 symmetry has been extensively studied \cite{23SYM}. It leads to maximal atmospheric
mixing with $\sin^2\theta_{23}=1$, and a vanishing reactor angle with $\sin^2\theta_{13}=0$. The mixing matrix is real, so no
Dirac phase is present, but the three eigen-masses are unknown and could be complex, thus admitting arbitrary 
Majorana phases and neutrino masses.

As we shall show, the HPS neutrino mass matrix is also magic, 
in the sense that the sum of each column and the sum of 
each row are all identical. I call such a matrix {\it magic} because it reminds me of magic squares, though the latter also
have identical diagonal sums.

It is this magic property of the mass matrix that we want to investigate, with or
without 2-3 symmetry. For specific examples, see, for example, Ref.~\cite{HS}.
It turns out that the mixing
matrix for magic mass matrices can be parameterized in the way proposed recently by Bjorken, Harrison, and Scott (BHS)
\cite{BHS}, with  arbitrary Majorana phases and neutrino masses.

\section{HPS Mixing Matrix}
The HPS form of the MNS \cite{MNS} mixing matrix is 
\be U_{HPS}=\pmatrix{2/\sqrt{6}&1/\sqrt{3}&0\cr -1/\sqrt{6}&1/\sqrt{3}&1/\sqrt{2}\cr -1/\sqrt{6}&1/\sqrt{3}&-1/\sqrt{2}\cr}.\ee
If $m_i$ are the neutrino masses, possibility  complex to absorb
the two Majorana phases, and if $m={\rm diag}(m_1,m_2,m_3)$, 
then the neutrino mass matrix may be written
\be
M_{HPS}=U^T_{HPS}mU_{HPS}={1\over 6}\pmatrix{4m_1+2m_2&-2m_1+2m_2&-2m_1+2m_2\cr
-2m_1+2m_2&m_1+2m_2+3m_3&m_1+2m_2-3m_3\cr
-2m_1+2m_2&m_1+2m_2-3m_3&m_1+2m_2+3m_3\cr}.\ee
It is 2-3 symmetric. It is also magic, because the row sums and column sums are each equal to $m_2$.
Conversely, we shall show in the next section that a magic and 2-3 symmetric matrix gives rise to the mixing structure
(1), so that a magic 2-3 symmetry for the mass matrix is synonymous to a HPS mixing structure.

\section{Magic Matrix}
An $n\times n$ matrix $A$ will be called {\it magic} if the row sums and the column sums are all equal to a common number 
$\alpha$:
\be
\sum_{i=1}^nA_{ij}=\sum_{j=1}^nA_{ij}=\alpha.\ee
For example, every permutation matrix of $n$ objects has a single 1 in every row and every column, and 0 elsewhere,
so it is a magic matrix with $\alpha=1$.

It is easy to see that magic matrices are closed under addition, multiplication, inverse, and  scalar multiplication. In
other words, if $c$ is a number, and $A,B$ are magic matrices with common sums
$\alpha,\beta$, then $A+B, AB, A^{-1}$, and $cA$ are all magic, with common sums $\alpha+\beta, \alpha\beta,
\alpha^{-1}$, and $c\alpha$. 
Such closure properties are also true for 2-3 symmetric matrices.

Eq.~(3) tells us that $A$ has an eigenvector $v$, with eigenvalue $\alpha$. Namely, if all the
$n$ components $v_i$ of the vector $v$ are equal, then 
\be \sum_jA_{ij}v_j=\alpha v_i,\qquad\sum_iv_iA_{ij}=\alpha v_j.\ee
This eigenvector is normalized if we choose $v_i=1/\sqrt{n}$.

We specialize now to $3\times 3$ matrix $M$ that is magic.
To satisfy (3), it must have the general form
\be 
M=\pmatrix{a&b&c\cr e&d&a+b+c-d-e\cr b+c-e&a+c-d&d+e-c\cr},\ee
so it is described by 5 complex parameters. If $M$ is symmetric, as Majorana neutrino mass matrices are, then $e=b$, and
we are left with 4 complex parameters. As we shall see, they are the three complex masses $m_i$ and $U_{e3}$.
If $M$ is 2-3 symmetric, then $b=c=e$, making $M$ automatically symmetric,
and we are left with only 3 parameters, the three complex masses. This is the case with the HPS mass matrix in (2).

Our task is to find  the mixing matrix $U$  for a symmetric magic
mass matrix $M$, {\it i.e.,} a unitary matrix $U$ so that $U^TMU$ 
becomes a
diagonal matrix $m={\rm diag}(m_1,m_2,m_3)$. Since $U$ is unitary, we can write this relation as $MU=U^*m$, so that
\be
Mu_i=m_iu_i^*\ee
if $u_1,u_2,u_3$ are the three column vectors of $U$.
Since the normalized eigenvector $v$ in (4) is real, it satisfies (6) and is one of the three column vectors $u_i$.
By comparing with the HPS mixing matrix (1), we see that $v=u_2$. In other words, the second neutrino mass
eigenstate $\nu_2$ always mixes democratically with all three neutrino flavor states, thus a magic mass matrix leads to
a trimaximal mixing for $\nu_2$.
This is the basic assumption involved in the Bjorken-Harrison-Scott (BHS) parameterization
\cite{BHS} of the mixing matrix. The rest of the matrix elements in $U$ is determined by unitarity and allowed phase
choices. The result, as given by BHS, is
\be
U=\pmatrix{2C/\sqrt{6}&1/\sqrt{3}&U_{e3}\cr  -C/\sqrt{6}-\sqrt{3}U^*_{e3}/2&1/\sqrt{3}&C/\sqrt{2}-U_{e3}/2\cr
-C/\sqrt{6}+\sqrt{3}U^*_{e3}/2&1/\sqrt{3}&-C/\sqrt{2}-U_{e3}/2\cr},\ee
where $U_{e3}$ is complex so it contains the Dirac phase, allowing CP violation, and $C=\sqrt{1-3|U_{e3}|^2/2}$.

Note that 
\be\sum_i U_{ij}=0\ee
for $j=1$ and 3, because these two columns must be orthogonal to the second column. 

If $M$ is 2-3 symmetric as well, then  $U_{e3}=0$ \cite{23SYM}, so $U=U_{HPS}$ as claimed.

Conversely, in terms of the neutrino masses $m_i$ (possibly complex after absorbing the Majorana phases),
the mass matrix $M=UmU^T$ can be calculated from (7). Such a matrix is necessarily magic on account of (8):
\be 
\sum_iM_{ij}=\sum_{i,k}U_{ik}m_kU_{jk}=\sum_iU_{i2}m_2U_{j2}=m_2=\sum_jM_{ij}.\ee

\section{Conclusion}
In conclusion, present experimental data are consistent with the HPS mixing 
of eq.~(1), which exhibits a trimaximal mixing for $\nu_2$ and a bimaximal mixing for
$\nu_3$. In the basis where the charged-lepton mass matrix is diagonal, the neutrino mass matrix $M$ 
possesses both magic symmetry and 2-3 symmetry.
If magic symmetry is broken but 2-3 symmetry is kept, then \cite{23SYM} the bimaximal structure of $\nu_3$ is retained, 
but the trimaximal nature of $\nu_2$ is broken, thus
keeping atmospheric mixing maximal and the reactor angle zero, but the solar mixing  is
left as a free parameter. If
 2-3 symmetry is broken but magic symmetry is kept,
then it retains the trimaximal structure of $\nu_2$ while breaking the bimaximal nature of $\nu_3$. 
The mixing matrix in this case is parameterized by Bjorken, Harrison,
and Scott \cite{BHS},
with a single complex parameter $U_{e3}$, thus allowing CP violation. There are 
also two relations between the mixing parameters,
arising from the trimaximal structure  $U_{e2}=U_{\mu 2}=U_{\tau 2}=1/\sqrt{3}$.
In terms of the usual Chau-Keung parameterization of the mixing angles \cite{CK}, one relation is
$\sin\theta_{12}\cos\theta_{13}=1/\sqrt{3}$, and the other  involves the Dirac phase angle.

If both the 2-3 and the  magic symmetries are violated, then we are back to the general case with a full blown
Chau-Yeung parameterization. 

This research is support by the Engineering and Natural Science Research Council of Canada.


\end{document}